\documentclass{mem}
\usepackage{natbib}
\usepackage{txfonts}
\usepackage{balance}
\usepackage{graphicx}
\idline{1}{1}
\begin{document}

\title{\textit{HST} Proper Motions of Stars within Globular Clusters}
   \subtitle{}
\author{
A.\ Bellini\inst{1},
R.~P. van der Marel\inst{1},
\and J.\ Anderson\inst{1}
       }
\offprints{A. Bellini}
   
\institute{Space Telescope Science Institute,
3700 San Martin Drive, Baltimore, MD, 21218, USA
\email{bellini@stsci.edu}}

\authorrunning{Bellini}

\titlerunning{\textit{HST} proper motions of globular clusters}

\abstract{The stable environment of space makes \textit{HST} an
  excellent astrometric tool.  Its diffraction-limited resolution
  allows it to distinguish and measure positions and fluxes for stars
  all the way to the center of most globular clusters.  Apart from
  small changes due to breathing, its PSFs and geometric distortion
  have been extremely stable over its 20-year lifetime.  There are now
  over 20 globular clusters for which there exist two or more
  well-separated epochs in the archive, spanning up to 10+ years.  Our
  photometric and astrometric techniques have allowed us to measure
  tens of thousands of stars per cluster within one arcmin from the
  center, with typical proper-motion errors of $\sim$0.02 mas/yr,
  which translates to $\sim$0.8 km/s for a typical cluster.  These
  high-quality measurements can be used to detect the possible
  presence of a central intermediate-mass black hole, and put
  constraints on its mass. In addition, they will provide a direct
  measurement of the cluster anisotropy and equipartition. We present
  preliminary results from this project, and discuss them in the
  context of what is already known from other techniques.
  \keywords{Astrometry: proper motions -- Stars: Population II --
    Galaxy: globular clusters }} \maketitle{}

\section{Introduction}
We measured high-precision proper motions (PMs) in the cores of 23
globular clusters (GCs) for which good \textit{HST} data exist. We
focused our attention on ACS/WFC, ACS/HRC, and WFC3/UVIS exposures,
which provide smaller pixel sizes, higher stability of the geometric
distortion solution, and higher dynamic range than other \textit{HST}
detectors.  The biggest step forward provided by our velocity
measurements will be the large number of PMs within the crowded core
of each cluster, from the tip of the red-giant branch down to faint
main-sequence stars.  The time baseline for these clusters goes from 2
up to 10 years.

Although the stable environment of space offers unsurpassed image
quality and stability, reaching high-precision astrometry (and
photometry) requires a careful treatment of possible systematic
effects: undersampling, space- and time-variable PSF models,
geometric-distortion corrections, charge-transfer inefficiency, and
transformations between dithered images. All these issues have been
properly addressed by our group, and we are working in further
improving our solutions (Anderson \& King 2006a, 2006b, Bellini \&
Bedin 2009; Anderson \& Bedin 2010; Bellini, Anderson \& Bedin 2011).

We consider each exposure as representing a stand-alone epoch, and
transformed star positions from each exposure onto the reference frame
(usually the GC treasury catalogs from Sarajedini et al.\ 2007). Then,
for each star we fit the two position components as a function of the
epoch, assuming linear motion. We iteratively rejected the most
extreme point if it is not in agreement with the trend to within its
expected error.

In the following, we will give an introductory view of the project,
anticipated scientific results, and some examples.

\section{The catalogs}

Table~1 lists analyzed clusters and the number of objects for which we
measured high-precision proper motions. The field-of-view varies from
cluster to cluster, depending on the detector used and the overlap
among epochs. The proper motion accuracy is of the order of 0.8 km/s
for well-exposed stars at a typical cluster distance, assuming an
average time baseline. These PM measurements are still preliminary,
and we are in the process of fine tuning our techniques to further
improve them and minimize possible source of systematics.

\begin{table}[t!]
\label{tab1}
\begin{center}
\footnotesize{
\begin{tabular}{crcr}
\multicolumn{4}{c}{{\textsc{Table 1}}}\\
\multicolumn{4}{c}{{\textsc{Analyzed Clusters and PM catalogs size}}}\\
\hline
\hline
\textbf{Cluster}&$\!\!\!\!\!$\textbf{$N$ sources}&
\textbf{Cluster}&$\!\!\!\!\!$\textbf{$N$ sources}\\
\hline
NGC 104  & 106$\,$000& NGC 6388& 37$\,$000 \\
NGC 288  & 15$\,$000 & NGC 6397& 14$\,$000 \\
NGC 362  & 74$\,$000 & NGC 6441& 57$\,$000 \\
NGC 1851 & 88$\,$000 & NGC 6535& 4000      \\
NGC 2808 & 17$\,$000 & NGC 6624& 2000      \\
NGC 5139& 293$\,$000 & NGC 6656& 54$\,$000 \\
NGC 5904& 71$\,$000  & NGC 6681& 28$\,$000 \\
NGC 5927& 63$\,$000  & NGC 6715& 74$\,$000 \\
NGC 6121& 9000       & NGC 6752& 47$\,$000 \\
NGC 6266& 57$\,$000  & NGC 7078& 76$\,$000 \\
NGC 6341& 84$\,$000  & NGC 7099& 2000      \\
NGC 6362& 8700  &&\\
\hline
\end{tabular}}
\end{center}
\end{table}

\begin{figure}[t!]
\includegraphics[width=\columnwidth]{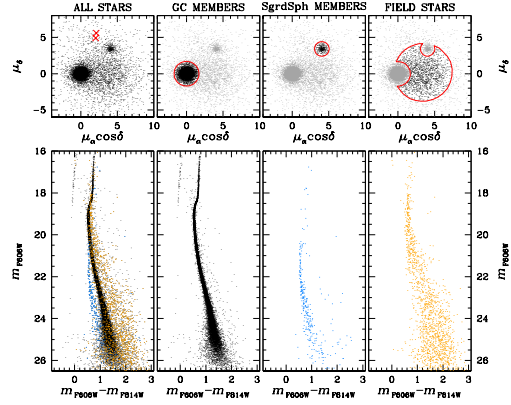}
\caption{\footnotesize \textit{Top panels}: proper-motion diagrams in
  the core of the GC NGC~6681. From left to right: all measured
  sources (red crosses mark the location of 2 galaxies in the field);
  selected cluster members, selected Sgr dSph members, and field
  stars. \textit{Bottom panels}: CMDs for all sources, GC members
  (black), Sgr dSph members (azure), and field stars (yellow).}
\label{f1}
\end{figure}

\begin{figure}[th!]
\includegraphics[width=\columnwidth]{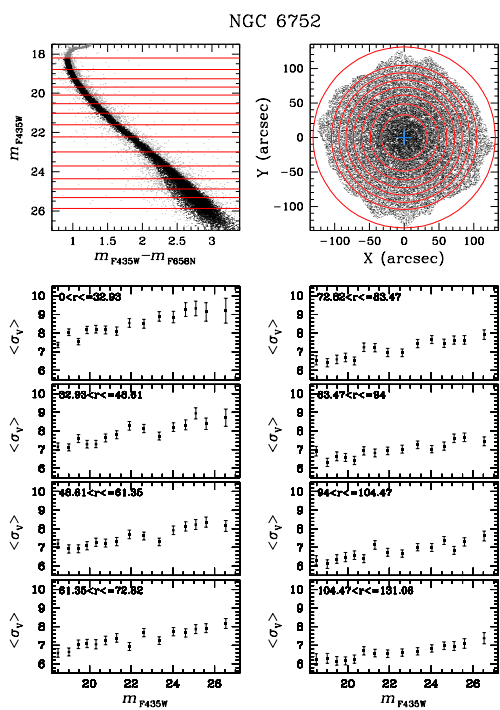}
\caption{\footnotesize Velocity-dispersion analysis for NGC~6752. Top
  panels show the selection of main-sequence stars in bins of
  magnitude (left) and radial distance from the cluster center
  (right). In the bottom panels, for each radial interval we plot the
  average velocity dispersion as a function of the magnitude. }
\label{f2}
\end{figure}

\begin{figure}[th!]
\includegraphics[width=\columnwidth,height=4cm]{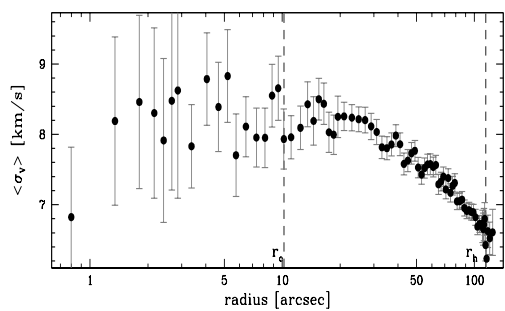}
\caption{\footnotesize $\langle \sigma_V \rangle$ as a function of the
  radial distance for all NGC~6752 stars in the field.}
\label{f3}
\end{figure}

\subsection{Anticipated science results}

High-precision proper motions in GCs allow us to undertake quite a
large variety of scientific investigations. To name a few:
\begin{itemize}
\item{\emph{cluster-field separation}: in order to study a bona-fide
  sample of cluster members;}
\item{\emph{internal motions}: to study kinematics and dynamics of GCs;}
\item{\emph{multiple-populations individual kinematics/dynamics}: to
  look for fossil signatures of a different formation process;}
\item{\emph{Absolute motions}: e.g., using background galaxies or
  field stars as a reference;}
\item{\emph{geometric distance}: by comparing the line-of-sight
  velocity dispersion with that on the plane of the sky;}
\item{\emph{GCs rotation on the plane of the sky}: from a measure of
  the stellar velocities as a function of position angle in different
  radial bins;}
\item{\emph{Equipartition}: from the analysis of stars' velocity
  dispersion as a function of the their mass;}
\item{\emph{radial mass distribution}: by studying stars' velocity
  dispersion as a function of their distance from the cluster center;}
\item{\emph{(an)isotropy}: by comparing tangential and radial components
  of stars' motion;}
\item{\emph{full 3D dynamics}: in case also line-of-sight
  velocities are known;}
\item{\emph{constraints on the presence of an intermediate-mass black
    hole}: looking both for fast-moving individual stars and for an
  increase in central velocity-dispersion-profile (van der Marel \&
  Anderson 2010);}
\item{etc\dots}
\end{itemize}

Below, we will show a few examples of what can be done with
high-precision proper motions.

\subsection{Absolute proper motions}

Figure~\ref{f1} shows proper-motion diagrams (on top) and the
corresponding CMDs (on bottom) for NGC~6681. It is clear from the
figure that we can: (i) clearly separate cluster members (in black)
from field stars (in yellow); (ii) isolate Sgr dSph stars (in azure)
in the background; and (iii) resolve the internal motion of the
cluster (because the stellar dispersion in the proper motion diagrams
is far larger than that of Sgr dSph stars, which in turn reflects our
internal proper motion errors).  We quickly identified by eye two
galaxies with sharp nuclei in our images, and marked their position on
the top-left panel, with red crosses. Their average location on the
proper-motion diagram can be used to define the zero-point of the
absolute motion of cluster, field (and in turn, constrain the Galaxy
potential), and Sgr dSph stars in the background.

\subsection{Mass segregation and equipartition}

Thanks to the extended field-of-view (up to $\sim$100 arcsec or more)
of most of the clusters (typically beyond the half-light radius), the
availability of high-precision PMs even in the dense clusters core,
and, most importantly, the fact that we can now measure kinematics as
a function of the position on the main sequence (stellar mass), for
the first time we have the possibility to directly study in detail the
degree of equipartition of each cluster (by measuring at the same time
stars' velocity and mass), using thousands of stars and two components
of the motion. 

Figure~2 illustrates the case for NGC~6752. The top panels show the
main sequence of the cluster (which we divided into 15
equally-populated magnitude bins, on the left), and the total
field-of-view (divided into 8 radial intervals, on the right). The
eight panels below show, for each radial interval, the average
velocity dispersion $\langle \sigma_V \rangle$ within each magnitude
bin.  We can easily see, from the figure, that inner stars move, on
average, faster than outer ones. Moreover, more massive stars are
slower than less massive ones.  In Figure~3 we extended our analysis
to all stars, regardless of their evolutionary stage, and plot their
$\langle \sigma_V \rangle$ as a function of the logarithmic radial
distance from the cluster center.  The distribution appears, within
the errors, to be flat inside $\sim$3 core radii, then it drops
steeply.

One way to infer whether a cluster has reached equipartition is to
measure the slope of the average velocity dispersion vs. stellar mass
in the log-log plane: a value of $-$0.5 would mean that the cluster is
in full equipartition. On the other hand, a flat slope implies no
equipartition at all. In Figure~4 we show this slope for a sample
of eight GCs.  The preliminary slopes we measured range from zero to
$-$0.5, and suggest that the paradigm that all GCs have reached
equipartition long ago may not be valid.

\section{Conclusions}

Our preliminary results clearly demonstrate the power of
high-precision proper motions of globular clusters. Many scientific
applications will benefit from this project.  Once our analysis has
been completed, we will make our proper-motion catalogs available to
the astronomical community, with photometry in the available filters.
Moreover, we will produce stacked images of the fields with accurate
WCS headers, so that users can easily cross-identify stars in our
catalogs with other surveys.

\begin{acknowledgements}
A.B., R.P.vdM, and J.A. acknowledge  support from STScI grants
 GO-11988, GO-12274, GO 12656, and GO-12845.
\end{acknowledgements}

\begin{figure}[t!]
\centering
\includegraphics[width=\columnwidth]{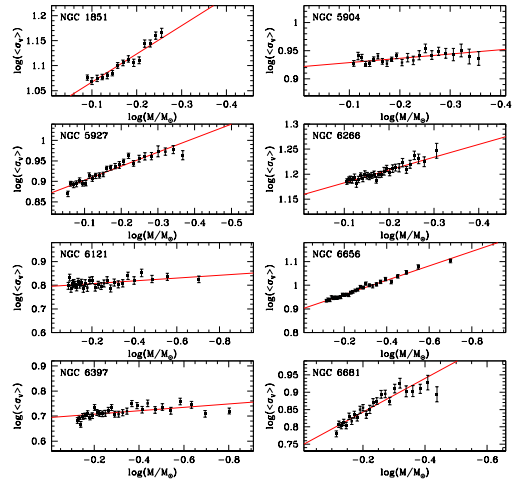}
\caption{\footnotesize Average velocity dispersion vs. Mass for a
  sample of 8 clusters, in the log-log plane. The scale ratio of the
  axes is the same in all panels. It is clear that the power-law
  relation between average velocity dispersion and stellar mass does
  not have a universal slope.}
\label{f4}
\end{figure}

\bibliographystyle{aa}

\end{document}